\documentclass[journal]{IEEEtran}
\usepackage[utf8]{inputenc}
\usepackage{graphicx}
\usepackage{url}
\usepackage{xcolor}
\usepackage[caption=false]{subfig}
\usepackage{amsmath,amssymb,amsfonts,flushend,tabularx,hyperref}
\usepackage{subfig}

\newcommand{\bparagraph}[1]{\par\noindent\textbf{#1}}
\begin{document}

\title{Reliable Detection of Compressed and Encrypted Data}

\author{Fabio De Gaspari, Dorjan Hitaj, Giulio Pagnotta, Lorenzo De Carli, Luigi V. Mancini
  \thanks{F. De Gaspari, D. Hitaj, G. Pagnotta and L.V. Mancini are with Dipartimento di Informatica, Sapienza Universit\`{a} di Roma, 00198 Rome, Italy (Email: \{degaspari, hitaj.d, pagnotta, mancini\}@di.uniroma1.it).}
  \thanks{L. De Carli is with the Department of Computer Science, Worcester Polytechnic Institute, Worcester, MA 01609, USA (email: ldecarli@wpi.edu).}}

\maketitle

\begin{abstract}
Several cybersecurity domains, such as ransomware detection, forensics and data analysis, require methods to reliably identify encrypted data fragments. Typically, current approaches employ statistics derived from byte-level distribution, such as entropy estimation, to identify encrypted fragments. However, modern content types use compression techniques which alter data distribution pushing it closer to the uniform distribution. The result is that current approaches exhibit unreliable encryption detection performance when compressed data appears in the dataset. Furthermore, proposed approaches are typically evaluated over few data types and fragment sizes, making it hard to assess their practical applicability. 

This paper compares existing statistical tests on a large, standardized dataset and shows that current approaches consistently fail to distinguish encrypted and compressed data on both small and large fragment sizes. 
We address these shortcomings and design \textsc{EnCoD}, a learning-based classifier which can reliably distinguish compressed and encrypted data. We evaluate \textsc{EnCoD} on a dataset of 16 different file types and fragment sizes ranging from 512B to 8KB. Our results highlight that \textsc{EnCoD} outperforms current approaches by a wide margin, with accuracy ranging from $\sim82\%$ for 512B fragments up to $\sim92\%$ for 8KB data fragments. Moreover, \textsc{EnCoD} can pinpoint the exact format of a given data fragment, rather than performing only binary classification like previous approaches.

\end{abstract}

\section{Introduction}
\label{sec:Introduction}

Reliable detection of encrypted data fragments is an important primitive with many applications to security and digital forensics. For instance, ransomware detection algorithms use estimates of write-operations' data randomness to quickly identify evidence of malicious encryption processes~\cite{kirda_redemption,mehnaz_rwguard,continella_shieldfs:_2016,kirda_unveil}. When performing digital forensic analysis of hard drives and phones, it is oftentimes important to identify encrypted archives~\cite{conti_automated_2010}. Finally, encryption detection is widely used in network protocol analysis~\cite{de_carli_botnet_2017,dorfinger_real-time_2010}. 

A popular approach to address this problem is to estimate the Shannon entropy of the sequence of interest using the Maximum Likelihood Estimator (MLE):  $\hat{H}_{MLE}$. This approach leverages the observation that the distribution of byte values in an encrypted stream closely follows a uniform distribution; therefore, high entropy is used as a proxy for randomness. This estimator has the advantage of being simple and computationally efficient. As non-encrypted digital data is assumed to have low byte-level entropy, the estimator is expected to easily differentiate non-encrypted and encrypted content.

While this approach remains widely used (e.g.,~\cite{kirda_redemption,mehnaz_rwguard,continella_shieldfs:_2016,kirda_unveil}), a number of works have highlighted its limitations. Modern applications tend to compress data prior to both storage and transmission. Popular examples include the zip compressed file format, and HTTP compression~\cite{http-rfc} (both using the DEFLATE algorithm). As compression removes recurring patterns in data, compressed streams tend to exhibit high Shannon entropy. As a result, compressed data exhibit values of $\hat{H}_{MLE}$ that are close and oftentimes overlapping with those obtained by encryption. In principle, compressed content can be identified by using appropriate parsers. However, many security-related applications, such as ransomware detection, traffic analysis and digital forensics, generally do not have access to whole-file information, but rather work at the level of \textit{fragments} of data. In these settings, the metadata that is required by parsers is not present or is incomplete~\cite{park_data_2008}.
Given this issue, a number of works have been looking at alternative tests to distinguish between encrypted and compressed content~\cite{malhotra_detection_2007,wang_using_2011,foresti_efficient_2016,lipmaa_data_2017,hahn2018detecting,casino_hedge_2019,choudhury_empirical_2020}. While these works have the potential to be useful, there has been limited evaluation of their performance on a standardized dataset. Consequently, there is no clear understanding of how these approaches: (i) fare on a variety of compressed file formats and sizes, and (ii) compare to each other. The potential negative implications are significant: the use of ineffective techniques for identifying encrypted content can hinder the effectiveness of ransomware detectors~\cite{gaspari2019}, and significantly limit the capability of forensic tools.

Our work compares state-of-the-art approaches on a large dataset of different data types and fragment sizes. We find that, while more useful than entropy estimates, current approaches fail to achieve consistently high accuracy. To address this, we propose \textsc{EnCoD} (\textbf{En}cryption/\textbf{Co}mpression \textbf{D}istinguisher), a novel neural network-based approach. Our evaluation shows that \textsc{EnCoD} outperforms existing approaches for most considered file types, over all considered fragment sizes. 

\textsc{EnCoD} can distinguish between compressed and encrypted data fragments as small as 512B with $86\%$accuracy.  The accuracy increases to up to $94\%$ when distinguishing between encrypted and purely compressed data (i.e.,zip, gzip), and up to $100\%$ in the case of compressed application data fragments (e.g., pdf, jpeg, mp3) when the fragment size is 8KB.
Furthermore, we investigate the applicability of robust feature extraction techniques such as autoencoders to our architecture, in an effort to understand whether feature vector pre-processing can lead to increased performance compared to a plain neural network (NN) architecture in this domain.

This paper revises and extends our previous conference paper~\cite{de2020encod}, by considering a larger and more diverse dataset of file fragments and evaluating the effectiveness of data pre-processing on accuracy. Overall, we make the following contributions:

\begin{itemize}
\item We review and categorize existing literature on the topic of distinguishing compressed and encrypted data fragments.
\item We build and make available to the community a large, standardized dataset of data fragments of different sizes from 16 different data formats.\footnote{The full dataset is available at~\url{https://drive.google.com/file/d/1IDNv3U1hRILXblwT9fI3G-D8hJquiequ}}
\item We systematically evaluate and compare state-of-the-art approaches on our dataset for different fragment formats and sizes.
\item We propose a new neural-network based approach and show that it outperforms current state-of-the-art tests in distinguishing encrypted from compressed content for most considered formats, over all considered fragment sizes.
\item We propose a new multi-class classifier that can label a fragment with high accuracy as encrypted data, general-purpose compressed data (zip/gzip/rar/bz2), or one of multiple application-specific compressed data (png, jpeg, pdf, mp3, office, video).
\item We investigate the effectiveness of data pre-processing techniques such as autoencoders for our architecture, and show that plain neural network models outperform these approaches in the considered domain.
\item We thoroughly discuss the implications of our findings and effectiveness of the evaluated approaches in distinguishing compressed and encrypted data.
\end{itemize}

The rest of this paper is structured as follows: Section~\ref{sec:Background} provides background on entropy estimation and its applications. Section~\ref{sec:Review} reviews existing approaches to the problem. Section~\ref{sec:Classifier} presents and evaluates a novel approach to the problem, based on deep learning. Section~\ref{sec:Evaluation} evaluates the performance of the considered approaches, discussing their strengths and limitations.  Section~\ref{sec:Discussion} discusses the implications of our findings. Section~\ref{sec:RelatedWork} discusses related work and Section~\ref{sec:Conclusions} concludes the paper.

\section{Background}
\label{sec:Background}

Determining the format of a particular data object (e.g. a file in permanent storage, or an HTTP object) is an extremely common operation. Under normal circumstances, it can be accomplished by looking at content metadata or by parsing the object. Things get more complicated, however, when no metadata is available and the data object is corrupted or partly missing. In this paper, we focus on detection of \textit{encrypted content} and, in particular, on distinguishing between encrypted and compressed data fragments.  We begin by examining relevant applications of encryption detection primitives.

\subsection{Ransomware Detection}

\textit{Ransomware} encrypts user files with the aim of making them unusable for the user. It then presents a prompt asking the user to pay a ransom in order to receive the decryption key. Ransomware attacks can cause significant financial damage to organizations~\cite{ransomware_atlanta:2018,ransomware_uk:2017,ransomware_trends_nyt}.

Mitigating a ransomware infection requires rapid detection and termination of all ransomware processes. 
A number of approaches based on \textit{behavioral process analysis} have been proposed for this purpose~\cite{kirda_redemption,mehnaz_rwguard,continella_shieldfs:_2016,kirda_unveil}. These approaches typically rely on a classifier trained on various process-related features to distinguish benign and ransomware processes. Virtually all proposed behavioral detectors use entropy of file write operations as one of the key features, based on the insight that frequently writing encrypted content is a characteristic behavioral fingerprint of ransomware. Entropy is typically estimated using $\hat{H}_{MLE}$. In several approaches entropy is estimated on the content of individual file writes~\cite{kirda_redemption,kirda_unveil,continella_shieldfs:_2016}, therefore the estimation procedure has only access to partial file fragments.
  
\subsection{Forensics}

Digital forensics oftentimes involves analysis of phone~\cite{walls_forensic_2011} or PC~\cite{park_data_2008} storage that has been corrupted, or uses an unknown format. Therefore, forensic techniques attempt to recover data of interest (contacts, pictures, etc.) by searching for blocks with recognizable structure. These techniques typically only have access to data fragments, rather than whole files.

Encrypted and compressed data represent a corner case, as they exhibit a complete lack of structure. Still, detecting such content may be important in data recovery operations (e.g., if sensitive data is known to have been encrypted). Distinguishing between compressed and encrypted blocks is notoriously difficult, and some forensic approaches label data as \textit{``compressed or encrypted''}, without attempting to pinpoint which one of the two it is~\cite{conti_automated_2010}.

\subsection{Network Traffic Analysis}

Network traffic analysis examines flows in/out of a network to identify security issues. Regulations (e.g. HIPAA in the U.S.) and best practices expect sensitive data to be encrypted in transit; therefore, entropy-based analyzers have been proposed to ensure that all traffic leaving a monitored network is encrypted~\cite{dorfinger_real-time_2010}. Another application is reverse-engineering of network protocols used by malware. It has been observed~\cite{de_carli_botnet_2017} that malware protocols may mix encrypted and non-encrypted content within the same message. Encryption detection primitives can be applied to break messages into encrypted and non-encrypted fields.

In both cases above, encryption detectors have partial visibility on the data stream and can only access fragments of data (e.g., an encrypted stream broken into individual packets), rather than whole data objects.

\begin{figure*}[t]
  \centering
  \includegraphics[width=.9\textwidth]{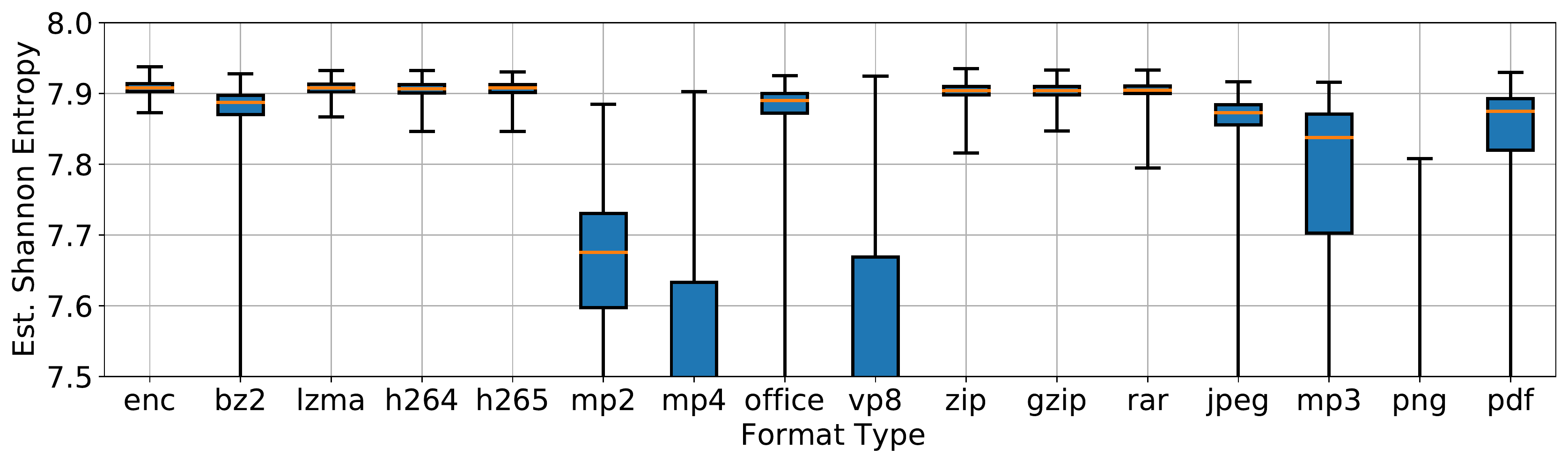}
  \caption{Entropy ranges for common formats (2048B blocks)}
  \label{fig:entro}
\end{figure*}

\subsection{Challenges}
\label{sec:Challenges}

In the three domains above, the use of Shannon entropy has been proposed in order to identify encrypted content. Entropy is used to measure the \textit{information content} of a byte sequence; highly structured data exhibit low entropy, while unstructured data---such as a randomly distributed sequence---have high entropy. Therefore, an entropy estimate can be used as a proxy for how close a sequence of bytes is to being randomly distributed. Most encryption algorithms output ciphertexts whose byte-value distributions tend to follow a uniform distribution. As a result, an encrypted bytestream will almost invariably exhibit high entropy.

One of the most common approaches to entropy estimation is the Maximum Likelihood Estimator $\hat{H}_{MLE} = -\sum_{i=0}^{255}{f_ilog_2(f_i)}$, where $f_i$ is the frequency of byte value $i$ in the sequence.
The entropy range is $[0-8]$. The frequency $f_i$ of byte value $i$, which is measurable, is used in place of the probability $P(i)$ of that value occurring, which is unknown.
This approach is commonly used in some of the applications above (e.g.,~\cite{kirda_redemption,dorfinger_real-time_2010}), due to its simplicity and efficiency.

This reasoning assumes that, while encrypted data has high entropy, non-encrypted data does not. This appears reasonable, as most relevant data types (e.g., text, images, audio) are information-rich and highly structured. However, this assumption does not hold true in modern computing. Modern CPUs can efficiently decompress data for processing, and compress it back for storage or transmission; this is oftentimes performed in real time and transparent to the user. As a result, most formats tend to apply compression~\cite{noauthor_docx_2017,wallace1992jpeg}. Informally, a good compression algorithm works by identifying and removing recognizable structures from the data stream; as a result, compressed data tend to exhibit high entropy. In practice, this fact compromises the ability of entropy-based detectors to distinguish encrypted and non-encrypted, compressed content.

\bparagraph{Entropy estimates for common data formats.} In order to substantiate the claim above, we computed entropy estimates using a dataset consisting of \textbf{10,000} file fragments. The dataset covers various popular file formats and AES-256-encrypted data. We considered multiple fragment sizes, from 512B to 8KB (details in Section~\ref{sec:Classifier}). Figure~\ref{fig:entro} summarizes the distribution of estimated entropy values for eight different formats with block size 2048 (some ranges truncated for clarity). Results for other block sizes were qualitatively similar. 
As illustrated in Figure~\ref{fig:entro}, both general-purpose (e.g., zip, rar) and domain-specific (e.g., jpeg, mp3) compression algorithms result in data which exhibits entropy whose ranges are overlapping with that of encrypted content (enc). The only format that can be unambiguously distinguished is png. Even so, png still overlaps with various other formats. Interestingly, utilities that create and modify data in zip, gzip, office and png format internally all use the DEFLATE algorithm for compression: the differences in entropy are likely due to differences in file structure and algorithm implementation.

Due to the limits of entropy estimation, the attention of the community has been increasingly focusing on alternative measures that can more precisely estimate whether data follow a random distribution. However, no comprehensive review of such approaches exists. In the next section, we review state-of-art approaches, while we evaluate and compare them in Section~\ref{sec:Evaluation}.

\section{Review of Existing Techniques}
\label{sec:Review}

This section reviews three state-of-the-art approaches to distinguishing encrypted and compressed content: the NIST suite, $\chi^2$ and HEDGE~\cite{casino_hedge_2019}. Strictly speaking, these approaches test the \textit{randomness} of a string of bytes, and make no attempt to determine its type. However, due to their high precision they can be used to distinguish true pseudorandom (encrypted) sequences and compressed ones which, while approximating a randomly generated stream, maintain structure. 

The NIST suite and $\chi^2$ are standard statistical tests for identifying randomly-distributed data. HEDGE is a recently proposed statistical approach which shows promising results. HEDGE is a combination of a subset of the NIST tests and two forms of $\chi^2$ tests. Note that, despite the inclusion of HEDGE, we decided to also report separate results for NIST and $\chi^2$ due to the fact that those are designed to be, and oftentimes are, used as standalone tests.

\subsection{NIST SP800-22}
\label{sec:ReviewNist}

The NIST SP800-22 specification~\cite{rukhin_statistical_2010} describes a suite of tests whose intended use is to evaluate the quality of random number generators. The suite consists of 15 distinct tests,  which analyze various structural aspects of a byte sequence. These tests are commonly employed as a benchmark for distinguishing compressed and encrypted content (e.g.,~\cite{casino_hedge_2019,choudhury_empirical_2020}). 
Each test analyzes a particular property of the sequence, and subsequently applies a test-specific decision rule to determine whether the result of the analysis suggests randomness or not. 
When using the NIST suite for discriminating random and non-random sequences, an important question concerns aggregation of the results of individual tests. Analysis of the tests~\cite{rukhin_statistical_2010} suggests that they are largely independent. Given this observation, and the intrinsic complexity of \textit{a priori} defining a ranking between the tests, we use a \textit{majority voting} approach. In other words, we consider a fragment to be random (and therefore encrypted) when the majority of tests considers it so. Since some of the tests require a block length much bigger than the ones we use for our smaller fragment sizes, we did not consider in the voting the tests that cannot be executed.

\subsection{$\chi^2$ Test}
\label{sec:ReviewChi2}

The $\chi^2$ test is a simple statistical test to measure goodness of fit. It has been widely applied to distinguish compressed and encrypted content~\cite{malhotra_detection_2007,lipmaa_data_2017,casino_hedge_2019}. Given a set of samples, it measures how well the distribution of such samples follows a given distribution. 
Mathematically, the test is defined as:

\vspace{0.1in}
$\chi^2 = \sum\limits_{i=0}^{255}\frac{(N_i - E_i)^2}{E_i}$
\vspace{0.1in}

\noindent where $N_i$ is the actual number of samples assuming value $i$, and $E_i$ is the expected number of samples assuming value $i$ according to the known distribution of interest. Since the distribution being evaluated for goodness of fit is the discrete uniform distribution, $\forall i E_i = L/256$, where $L$ is the particular fragment length being considered. The results of the test can be interpreted using either a fixed threshold, or a confidence interval~\cite{casino_hedge_2019}.

\subsection{HEDGE}
\label{sec:hedgepaper}

HEDGE~\cite{casino_hedge_2019} simultaneously incorporates three methods to distinguish between compressed and encrypted fragments: $\chi^2$ test with absolute value, $\chi^2$ with confidence interval and a subset of NIST SP800-22 test suite. Out of the NIST SP800-22 test suite HEDGE incorporates 3 tests: \textit{frequency within block test}, \textit{cumulative sums test}, and \textit{approximate entropy test}. These tests were selected due to (i) their ability to operate on short byte sequences, and (ii) their reliable performance on a large and representative dataset. In the HEDGE detector the threshold of the number of the above-mentioned NIST SP800-22 tests failed is set to 0.
For the $\chi^2$ with absolute value test, the thresholds are pre-computed for each of the considered packet sizes, by considering the average and its standard deviation.
For $\chi^2$ with confidence interval, the $\chi\%$ interval is $(\chi\% > 99\% || \chi\% < 1\%)$.
For classifying the content of a packet, HEDGE applies the three randomness tests to the input data. Data is considered random only if it passes all tests.

\section{{\mdseries\textsc{EnCoD}}: A Learning-based Approach}
\label{sec:Classifier}

Past work and our own evaluation suggest that tests based on byte-value distribution, such as $\chi^2$, can distinguish some encrypted and compressed content, but have accuracy issues (ref. Section~\ref{sec:Evaluation}). Such tests, in a sense, ``collapse'' the entire distribution to a single scalar value, losing information concerning the shape of the distribution. It is therefore natural to ask if Deep Neural Networks (DNNs) can improve such results. DNNs can consider the entire discrete distribution (modeled as a feature vector), and can learn to recognize complex distributions~\cite{Lee0MRA17}.

 In order to evaluate the potential of DNNs we designed \textsc{EnCoD}, a set of two distinct neural network-based approaches for distinguishing encryption and compression.

\subsection{Model Architecture \#1: Binary Classifiers}
\label{sec:mabc}

Our first model is a binary classifier trained to distinguish a single specific compressed format from encrypted content. It may be used in cases where only one compressed format is known to exist in the dataset (e.g., detecting writes of encrypted data performed by a potential ransomware on image files vs legitimate writes of JPEG-compressed data).
We explored several alternative architectures for this application, and we found that the structure depicted in Figure~\ref{fig:binary_model} provides the best performance. The binary-classifier architecture consists of 4 fully-connected layers with dimensions as shown in the figure. We initialize the model weights using Glorot uniform~\cite{Glorot2010UnderstandingTD}. The activation function is ReLU for the first 3 layers, followed by a softmax on the output layer. We used a batch size of $64$ for training our model. Each hyperparameter has been chosen using grid search. We used the same procedure also for the model described in Section~\ref{sec:contenttypedet}.

\subsection{Model Architecture \#2: Content-Type Detector}
\label{sec:contenttypedet}

\begin{figure*}[t]
  \centering
  \subfloat[Binary Classifier Architecture\label{fig:binary_model}] {
     \includegraphics[width=.8\linewidth]{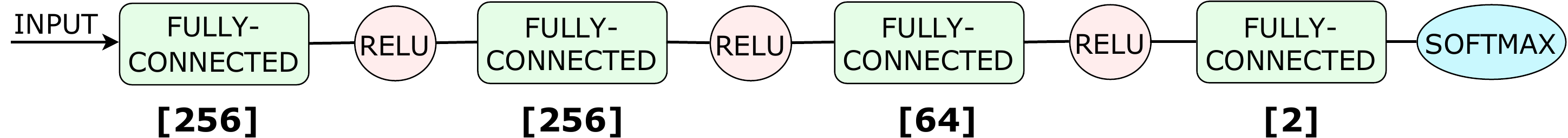}
  }
  \hfill
  \subfloat[Multi-Class Classifier Architecture\label{fig:multi_model}]{
     \includegraphics[width=1\linewidth]{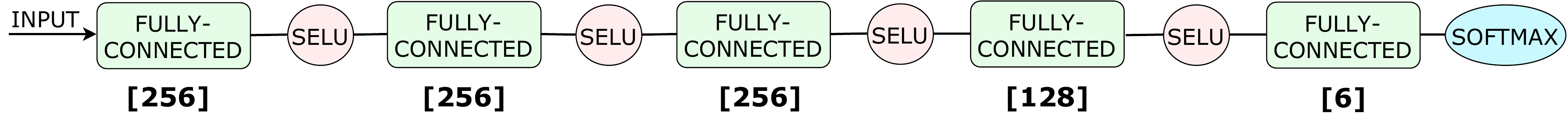}
  }
  \caption{Neural Network Architectures}
  \label{fig:model_arch}
 \end{figure*}

In many applications, a classifier may encounter more than one type of compressed data. Furthermore, it may be important to determine the specific type being encountered. To support these use cases, we design a \textit{content-type detector}: a multi-class classifier that can determine whether a given fragment is encrypted, or belongs to one of multiple known compressed formats. We explored several designs for the neural network, converging to the model depicted in Figure~\ref{fig:multi_model}. Its architecture consists of 5 fully-connected layers with dimensions as shown in the figure. We initialize model weights using LeCun normal~\cite{LeCun1998EfficientB}. Differently from the binary models, this multi-class classifier seemed prone to the dying neuron problem associated with the ReLU activation function~\cite{8260635}. We therefore opted for the SelU activation function~\cite{DBLP:journals/corr/KlambauerUMH17} for the first 4 layers, followed by a softmax on the output layer. We used a batch size of $64$ instances for training.

\subsection{Model Architecture Extension: Autoencoders}
\label{sec:autoencoder}
In addition to the architectures described in Section~\ref{sec:mabc} and Section~\ref{sec:contenttypedet}, we design a third model architecture that makes use of autoencoders. Autoencoders (AE) are NN models that are trained to compress an N-dimensional feature vector into an M-dimensional latent representation, with $M<N$, and then reconstruct the original input from the latent representation. AE are composed of two parts: the encoder, which compresses the original feature vector into the latent representation, and the decoder, which takes the output of the encoder and reconstructs the input. AE are generally used to extract robust features from a feature vector and aid classification~\cite{malphase,qi2014robust}. 

In our third architecture, we use the encoder portion of a trained autoencoder to pre-process samples into a compressed latent representation, which is then used as input by a fully-connected NN. We design two different set of NN with two different encoders:

\begin{itemize}
    \item AE architercure \#1: The encoder is composed of 3 fully-connected layers of size $256-200-128$ respectively, followed by a 3-layer fully-connected NN of size $128-64-2$.
    \item AE architecture \#2: The encoder is composed of 4 fully-connected layers of size $256-156-128-64$ respectively, followed by a 3-layer fully-connected NN of size $64-64-2$.
\end{itemize}

Both AE are trained for 25 epochs using the \textit{Adam} optimizer and \textit{Mean Squared error} as loss function. We initialize the weights using Glorot uniform for the NN model and uniform distribution for the AE. The activation function used is ReLU for all layers except for the output ones, which use softmax in the NN and sigmoid in the AE. We used a batch size of $64$ for training the NN and 128 for the AE. The hyperparameters were chosen using grid search.

\subsection{Fragment Dataset}
\label{sec:file_dataset}

We built a dataset of 400 million encrypted and compressed fragments from 16 different data formats. For the compressed data, we selected a set of formats covering common, popular content types. To generate the encrypted data fragments, we used the AES cipher in CBC mode implemented by the PyCryptodome library~\cite{pycrypto}. We chose AES because it is the most widely used and well known symmetric cipher, representative of modern ciphers which result in byte streams consistently close to random.

In constructing the dataset, we focused on ensuring a diversity of compressed formats, rather than compression algorithms. While algorithms such as DEFLATE are used in multiple compression formats, they are generally used with different parameters and/or embed compressed data in different ways within the compressed archive. Consequently, compressed archives created with different formats tend to differ considerably from each other even when using the same underlying compression algorithm.
This observation is empirically confirmed by our evaluation in Section~\ref{sec:Evaluation}. Finally, our dataset does not include data which is both compressed and encrypted, and we ensured such data is not present in the dataset. The dataset is comprised of the following data types:

\begin{enumerate}
    \item \textbf{AES encrypted data (enc)}. We used the AES implementation provided by the Cryptodome Python library. AES was configured to use CBC mode with 256-bit keys, with a random IV generated before encrypting each file.
\item \textbf{zip, gzip, rar, bz2, xz:}
DEFLATE, rar, Burrows–Wheeler and  Lempel–Ziv–Markov compressed data. These algorithms are among the most used for generic file compression, with  DEFLATE and rar being akin to de-facto standards. DEFLATE is also widely used for documents (such as in the Microsoft office file formats), and network applications (e.g., HTTP header compression).
\item \textbf{png and jpeg images:} png is used for lossless image compression; it internally uses DEFLATE, but png files present a structure that is different from that of zip files. jpeg uses DCT-based lossy compression.
\item \textbf{mp3 audio files:} MP3 compressors use a psychoacoustic model to remove inaudible frequencies from audio data, and compress the resulting data using a lossy algorithm based on the modified-DCT transform.
\item \textbf{pdf documents:} PDF is an office format used for document exchange and form filling. Internally, PDF files consist of a tree of objects that can be compressed using a variety of techniques. In practice, most PDF documents contain a large amount of compressed content, such as embedded images.
\item \textbf{Microsoft office files:} Ms office is one of the most used tool suites for office productivity. Internally, office files use the deflate algorithm for compression.
\item \textbf{h264, h265, mpeg2, mpeg4 and vp8 video formats:} h264 is one of the most widely used video codecs today, being the recommended codec for Youtube videos. H265 is the successor of h264 and substantially improves compression rate while maintaining the same video quality. Mpeg4 (Xvid codec) is a format that was widely used before h264. Vp8 and mpeg-2 are fairly dated video codecs that are not often used anymore, but there still exists old contents using them.
\end{enumerate}

\subsubsection{Fragment generation process.}

We generate fragments from a dataset of files:

\begin{itemize}
\item \textbf{zip/gzip/rar/bz2/xz/enc:} we used various textual documents obtained from a 2020 English Wikipedia dump~\cite{wiki_dump}. We created four copies of each file, each of which was either compressed using one of zip, gzip, rar, bz2, xz utilities (with default parameters), or encrypted using AES-256.
\item \textbf{png:} we crawled $\sim116,000$ png images from the web and various repositories~\cite{png_images}.
\item \textbf{jpeg: } we downloaded $\sim68,000$ images from the Open Images Dataset v5~\cite{jpeg_images} and various online sources.
\item \textbf{mp3:} we used the FMA medium dataset~\cite{mp3_files}, which contains 25,000 mp3 files.
\item \textbf{pdf:} we crawled $\sim3,000$ randomly-selected papers from arXiv~\cite{arxiv_2019}.
\item \textbf{office:} we sampled 4500 Word, 1700 PowerPoint and 1800 Excel files from a private hard drive. For privacy reasons, these files are not included in the provided dataset.
\item \textbf{video files:} we downloaded a large, h264-encoded video from Youtube and re-encoded it to the remaining formats.

\end{itemize}

We split each file into fragments of 512B, 1KB, 2KB, 4KB, and 8KB. In an effort to ensure that the dataset remains balanced, we randomly sampled 1M fragments for each fragment size/data type combination.

\subsection{Dataset Analysis Methodology}

\paragraph{Statistical tests (NIST, $\chi^2$, HEDGE)} For each fragment size, we randomly selected 10,000 compressed fragments (evenly distributed across the different compressed data types) and 10,000 encrypted fragments. We then executed the tests directly on these fragments.

\paragraph{\textsc{EnCoD}/Binary Classifiers}

We separately trained and evaluated classifiers for each fragment size. The features that are fed to our models for training/classification are derived from the histograms of the byte values for the observed fragment size. Each feature is the value of the probability density function at a given bin, normalized such that the integral over the range is 1.

We trained the binary classifiers by randomly selecting 3M vectors from the encrypted class and 3M vectors from the data type that we aim to distinguish. We partitioned this dataset into 85\% training, 5\% development and 10\% test. Before fitting the data to the model for training, we applied a MinMax scaler to scale the dataset from the range $[0,1]$ to the range $[0,2]$ (range selected via grid search). Scaling helps the ML model to more easily capture minute differences in the inputs, allowing to better distinguish among the classes and converge faster. 

\paragraph{\textsc{EnCoD}/Content-Type Detector}

To train the content-type detectors, for each fragment size we randomly sampled 6M feature vectors consisting of a mix of the considered file types. This dataset was partitioned into training, development and test sets in the same ratios used for the binary classifiers. We also scaled the dataset using the MinMax scaler with the same parameters used above.

\paragraph{\textsc{EnCoD}/Binary Classifiers with Encoder}

We trained one autoencoder per fragment size for all file types. The autoencoder was trained by randomly sampling 8M feature vectors in total (500,000 per each of the 16 file types). The feature vectors are created following the procedure used to train the plain binary classifiers. The dataset was partitioned into training, development and test sets following the same ratios used for the binary classifiers. 
The binary models are trained as previously described. The only variation is that the latent representation generated by the trained encoder is used as input by the NN, rather then the original feature vector derived from a given sample.

\section{Evaluation }
\label{sec:Evaluation}

This section comprehensively evaluate existing approaches (see Section~\ref{sec:Review}) and our own neural network-based approach, \textsc{EnCoD}. We frame the evaluation in terms of the following comparisons:

\begin{enumerate}
\item \textbf{Binary classification: all formats.} In Section~\ref{sec:bcaf}, we consider the ability of different detectors to discriminate encrypted and compressed data, regardless of the specific compressed format. Results show that our classifier outperforms NIST, $\chi^2$-test and HEDGE for all fragment sizes, with NIST performance approaching that of \textsc{EnCoD} only for large fragment size.
\item \textbf{Binary classification by format.} In Section~\ref{sec:bcbf}, we break down the performance of $\chi^2$, NIST and HEDGE by compressed format. We also report the performance of our per-format binary classifiers (see Section~\ref{sec:mabc}). The latter perform comparably or better than other tests on all formats but one.
\item \textbf{Format fingerprinting.} In Section~\ref{sec:ff}, we evaluate the accuracy of our multi-class classifier in labeling unknown fragments as the correct compressed format (or as encrypted). Results show that our classifier is able to distinguish the file type with an overall accuracy of 83\% for the 2048 byte fragment size. It also achieves high precision, especially on png, jpeg, mp3.
\item \textbf{Autoencoder approach.} In Section~\ref{sec:autoenc_eval} we analyze the performance of the encoder-based feature extraction approach in the binary classification task. Our evaluation shows that the plain NN approach outperforms the encoder-based consistently for all considered formats, with the exception of the pdf file type.
\end{enumerate}

\begin{figure*}[t]
  \centering
  \includegraphics[width=\textwidth]{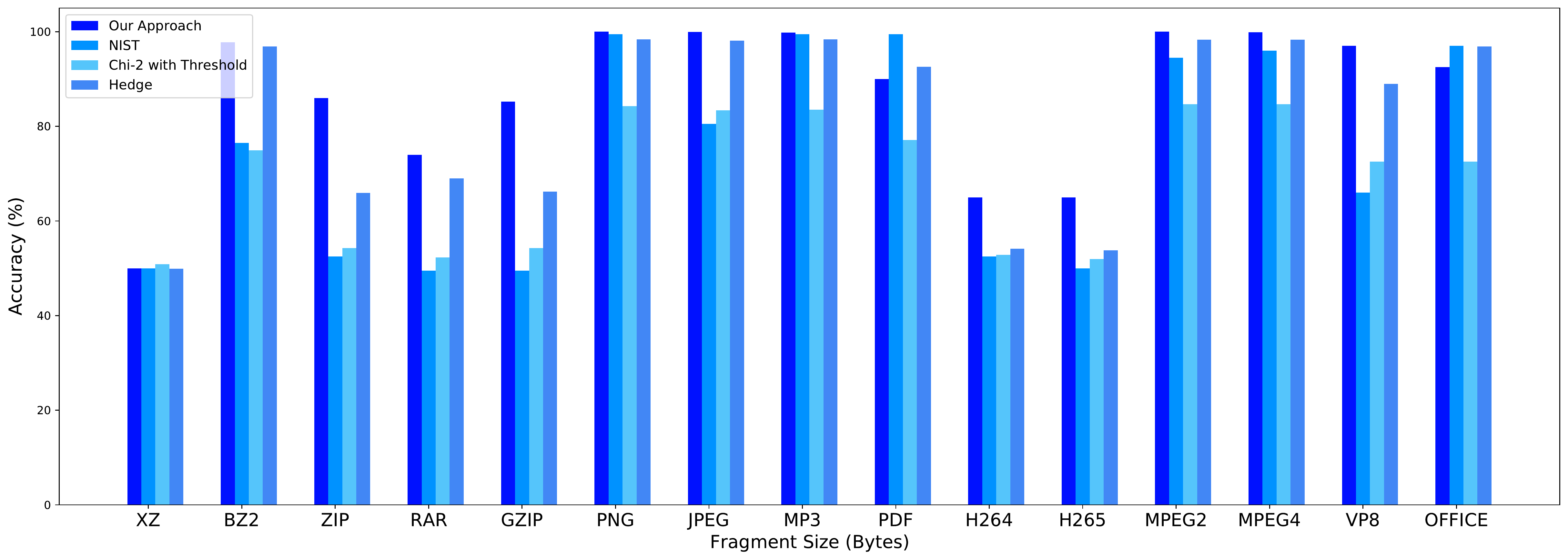}
  \caption{Performance comparison between our binary-classifier approach, NIST, $\chi^2$ with Threshold and HEDGE. All results are for 2048 Bytes data fragments.}
  \label{fig:bcbf}
\end{figure*}

\subsection{Implementation}

We implemented the classifier described in Section~\ref{sec:Classifier} using the Keras Library~\cite{chollet2015keras} for machine learning.  For the NIST tests, we used the official implementation~\cite{computer_security_division_nist_2016}. In order to aggregate the NIST tests results, we use the majority voting approach described in Section~\ref{sec:ReviewNist}. In order to label fragments as compressed or encrypted based on $\chi^2$ results, we used the thresholds suggested in the HEDGE paper~\cite{casino_hedge_2019}, as the analysis in HEDGE is specifically aimed at producing a dataset-independent threshold for general use. We implemented HEDGE according to the published description~\cite{casino_hedge_2019}. Finally, all experiments were conducted using the dataset described in Section~\ref{sec:Classifier}.
\begin{figure}[t]
  \centering
  \includegraphics[width=\linewidth]{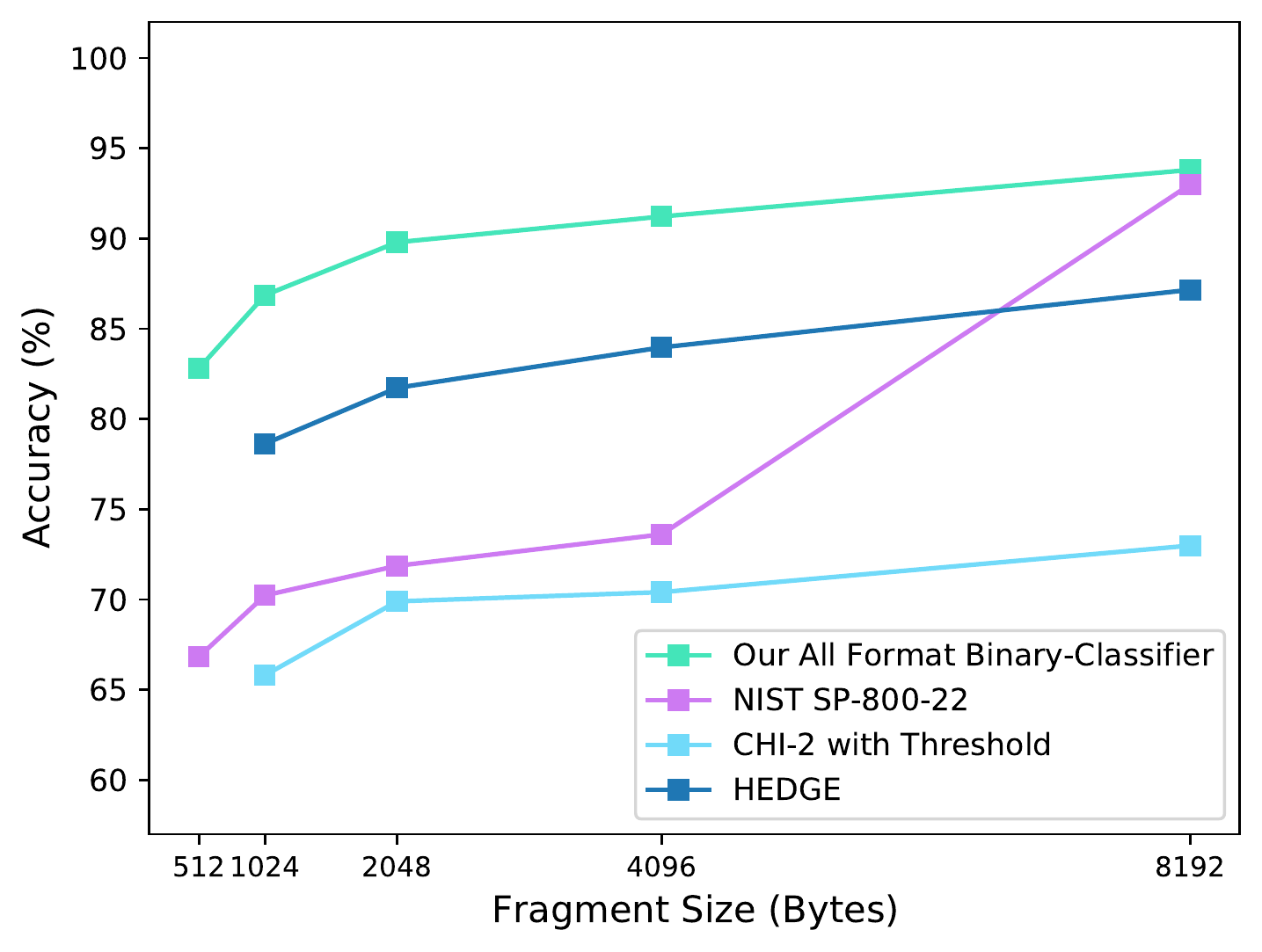}
  \caption{Performance comparison (binary classification: all formats) }
  \label{fig:bcaf}
\end{figure}

\subsection{Binary Classification: All Formats}
\label{sec:bcaf}

The first part of our evaluation considers the binary classification problem of determining whether a given high-entropy data fragment is compressed or encrypted. Given a fragment, the $\chi^2$ test, HEDGE, and the NIST test suite return whether the fragment's content appears random or not.
Therefore, a binary classifier can be constructed from the above-mentioned three tests by simply labeling the random content as encrypted. Our binary classifier used for this evaluation is based on our multi-class classifier. The multi-class classifier labels each fragment either as encrypted, or as one of the fourteen supported compressed formats. Since in this experiment we are only interested in distinguishing encryption and compression, regardless of the type, we combine all compressed type labels into one. Effectively, we consider classification in two labels: (1) a macro-label ``compressed'', which is comprised of the labels $\{$\emph{zip, rar, gzip, bz2, png, jpeg, mp3, pdf, h264, h265, mpeg2, mpeg4, vp8, office}$\}$ and (2) the label ``encrypted''. We analyze file type fingerprinting accuracy separately in Section~\ref{sec:ff}. It is worth noting that in experiment shown in Figure~\ref{fig:bcaf} we do not consider the xz file type. This is due to the fact that none of the considered approaches is able to reliably distinguish xz files from encryption for small fragment sizes, as we will see in Section~\ref{sec:bcbf}.

The results of this evaluation are depicted in Figure~\ref{fig:bcaf}. As we can see, the performance of all classifiers tends to improve as fragment size increases. This behavior is expected, as it is hard to approximate distribution information from short fragment sizes. However, as the fragment size increases, differences in distribution become more apparent and the models can exploit additional information for classification. We further discuss this phenomenon in Section~\ref{sec:Discussion}. In the binary classification task, \textsc{EnCoD}  outperforms all the other approaches on all block sizes, with the NIST approach reaching similar performance for $8K$ fragment size only. The $\chi^2$ accuracy remains consistently low across the range of block sizes, while performance for HEDGE increases for larger sizes but remains approximately $\sim10$ percentage points lower than \textsc{EnCoD}'s. These results suggest that the $\chi^2$ test has an intrinsic difficulty in discriminating non-random content which closely approaches a uniform random distribution.

\begin{figure*}[t]
\subfloat[Performance of our binary classifiers for compressed formats\label{fig:bin_all_comp}]{\includegraphics[width=.32\textwidth]{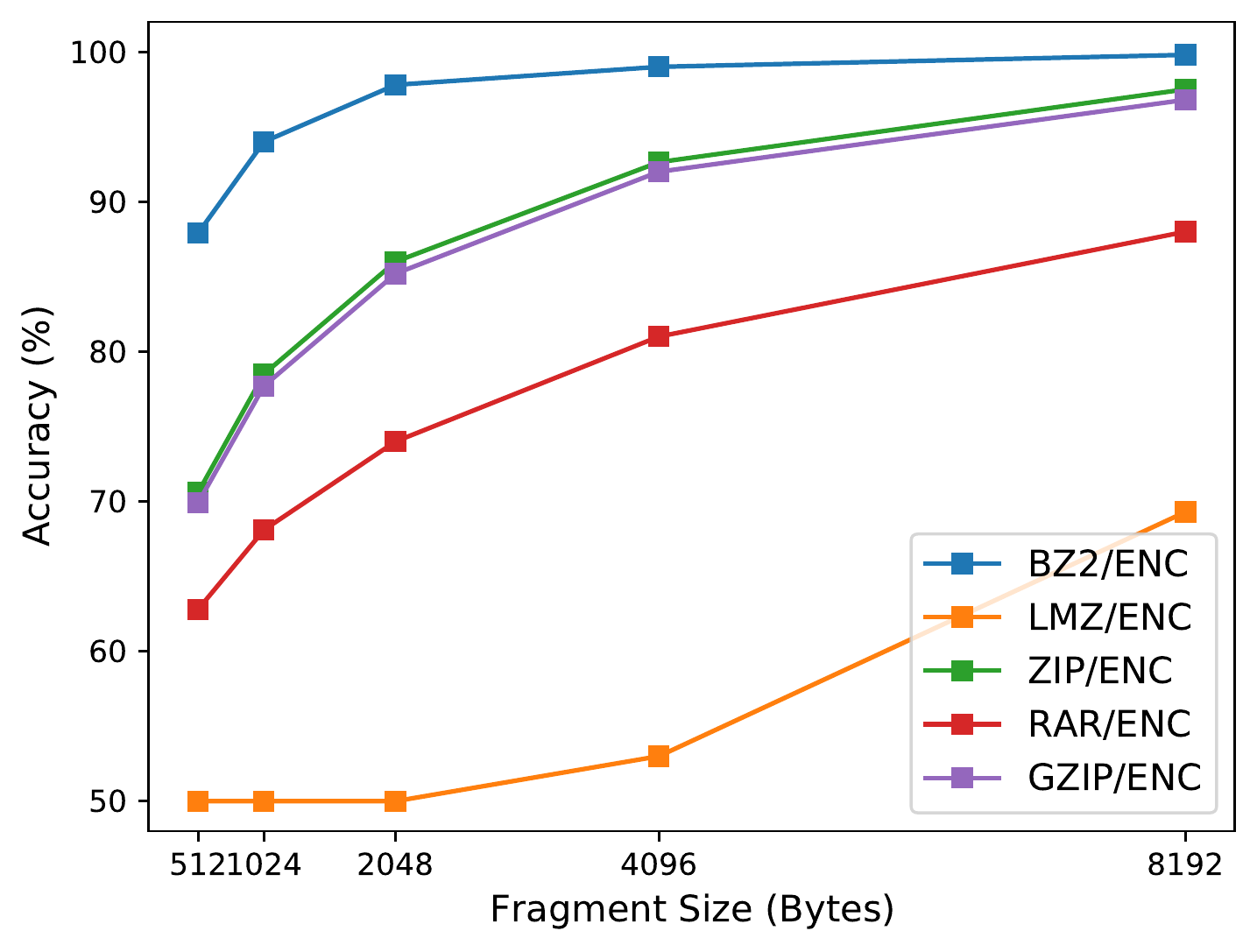}}\hfill
\subfloat[Performance of our binary classifiers for video formats \label{fig:bin_all_vid}]{\includegraphics[width=.32\textwidth]{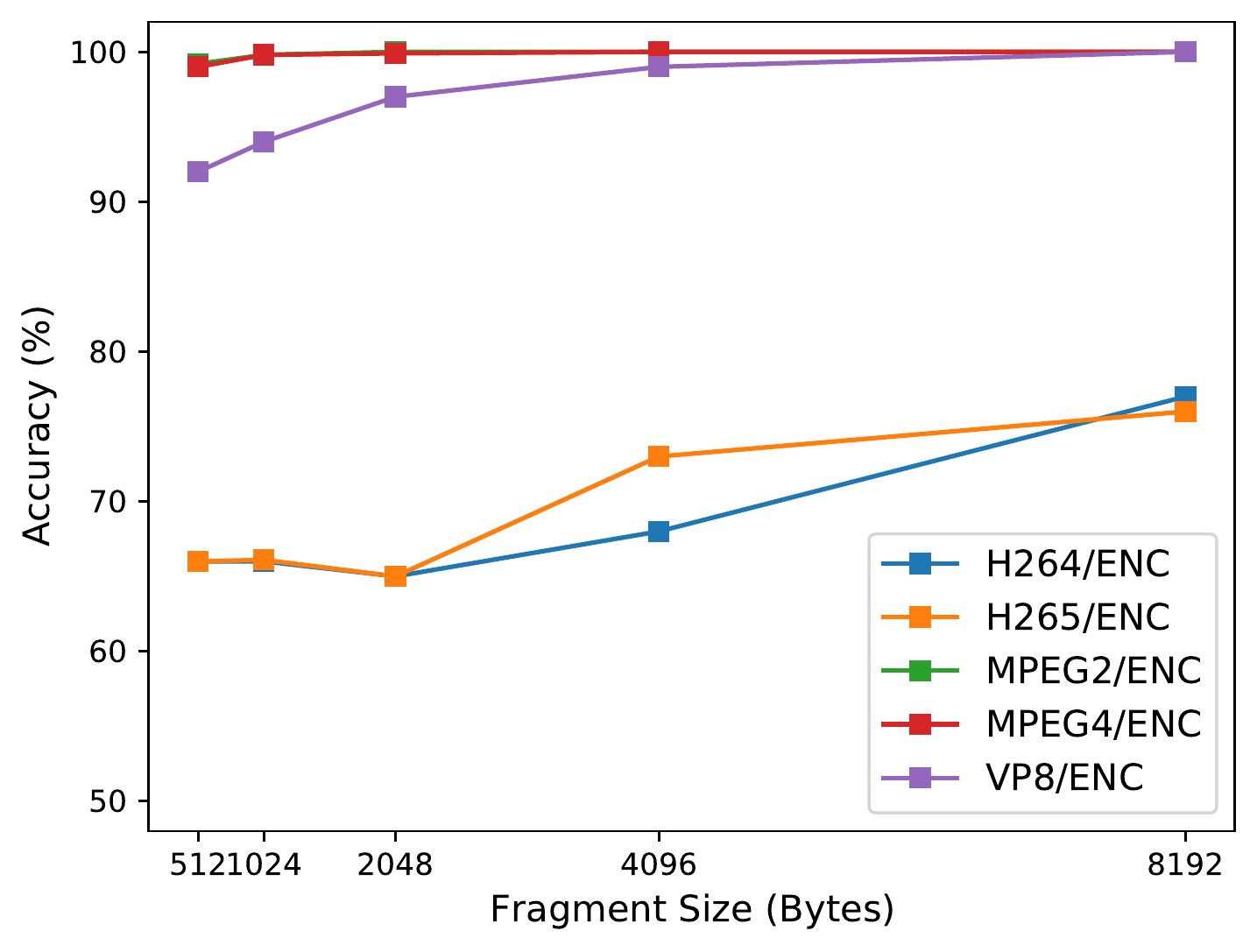}}\hfill
\subfloat[Performance of our binary classifiers for miscellaneous formats \label{fig:bin_all_misc}]{\includegraphics[width=.32\textwidth]{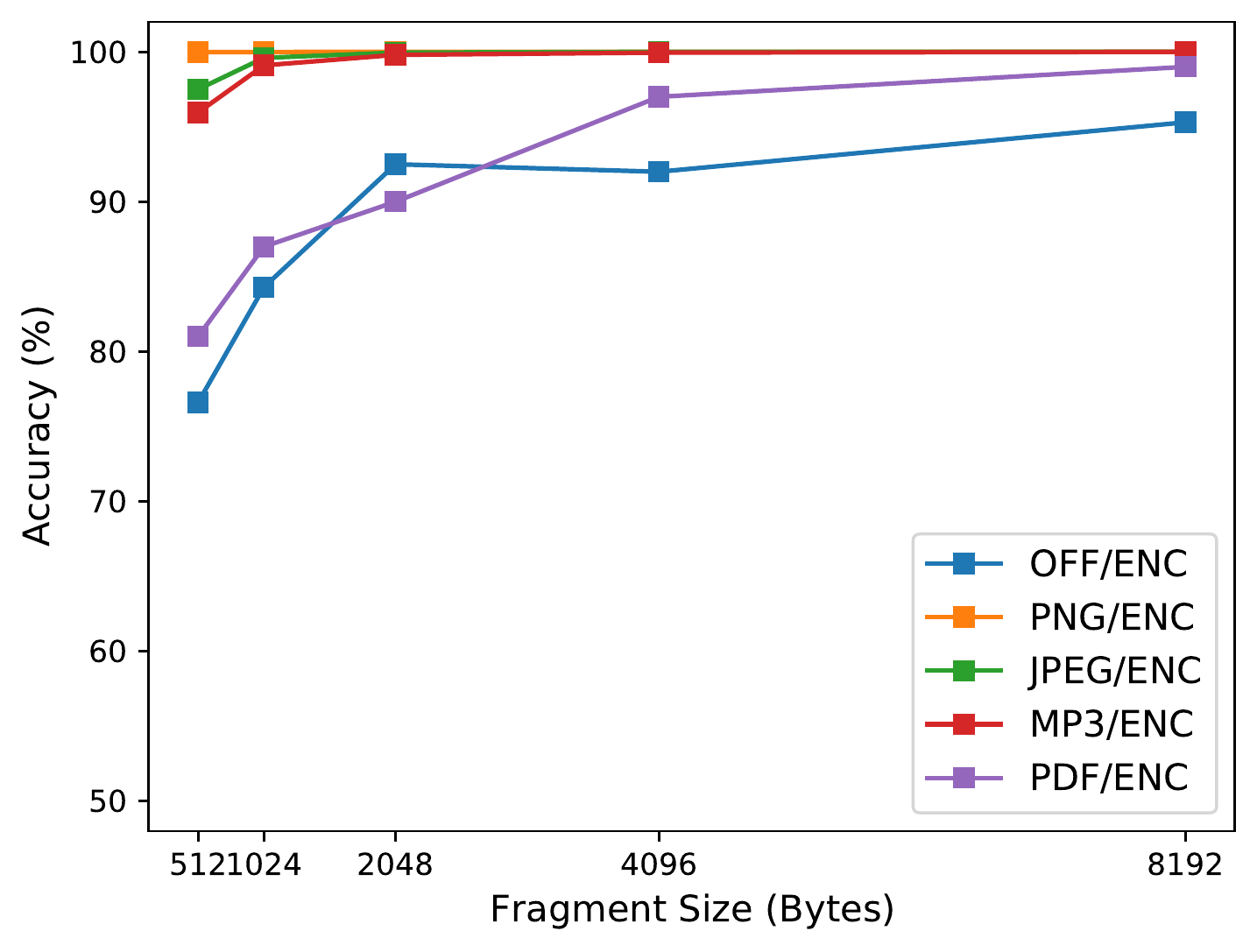}}\hfill
\caption{Performance of our binary classification models on different formats and block size ranging from 512 bytes to 8192 bytes.}
\label{fig:fig:binary_all_perf}
\end{figure*}

\subsection{Binary classification by format}
\label{sec:bcbf}

Our second experiment considers the question of whether some compressed formats are harder than others to distinguish from encrypted content. Such phenomenon may arise due to (i) differences in effectiveness between compression algorithms in removing redundancy (and thus structure) from the uncompressed data; and (ii) presence (or absence) of metadata, or other structured information interleaved with compressed data.

In order to answer this question, we break down results for the $\chi^2$-test, NIST suite, HEDGE test and \textsc{EnCoD} by format. In this experiment, for \textsc{EnCoD} we evaluate multiple binary NN classifiers, one per file type (see Section~\ref{sec:mabc}). Each type-specific classifier is trained to distinguish content of a given type from encrypted content (e.g., zip vs encrypted). Note that, while each of these classifiers is trained specifically on one format, the other tests ($\chi^2$, NIST and HEDGE) work the same regardless of the format. Despite this limitation, we believe this experiment to provide an informative analysis, as there are scenarios in which file type is known a priori, and we are interested only in differentiating between that type and encryption. For instance, if we consider a storage dedicated only to pictures backups, we can use a binary classifier (e.g., png/enc) to detect potential ransomware activity encrypting the pictures.

Figure~\ref{fig:bcbf} shows the comparison between the four approaches on 2048-byte blocks. Overall, neural network-based classifiers tend to fare better than the other tests, particularly on challenging formats such as zip/gzip, rar and bz2. PDF is the only format on which the NIST and HEDGE tests  outperform the neural network classifier, while for the office format \textsc{EnCoD} performs slightly worse but comparably. Interestingly, the $\chi^2$ fares slightly better than NIST on most formats, but its accuracy is significantly worse on formats that are typically easy to distinguish, such as PNG and MP3. We believe this to be due to the fact that the NIST tests look at a richer set of properties beyond byte value distribution, such as a presence of runs and repeated sequences. HEDGE test outperform $\chi^2$ on all file types, while outperforming NIST on most formats, beside PDF, and have similar performance on PNG, MP3, office and all video formats but vp8. Intersingly, all approaches fail to consistently distinguish the xz compressed type (Lempel–Ziv–Markov chain algorithm), with accuracy at $\sim50\%$. This behavior is indicative of the great compression performance of this algorithm, which pushes compressed data extremely close to a uniform distribution. Due to the inability of any of the considered approaches to consistently detect xz files, this type is not included in any of the aggregated results presented in this evaluation.

Finally, Figure~\ref{fig:fig:binary_all_perf} presents the performance of all our binary classifiers across all considered fragment sizes. The results are split in three images for clarity, all images have the same scale. These results highlight once again how accuracy increases significantly as block size increases. Indeed, for larger fragment sizes \textsc{EnCoD} can successfully classify xz compressed files from encryption reliably, with approximately $70\%$ accuracy on 8k fragment size, compared to $50\%$ for fragment sizes in the range $[512-2k]$.

\subsection{Format Fingerprinting}
\label{sec:ff}
\begin{figure}[t]
  \centering
  \includegraphics[width=\linewidth]{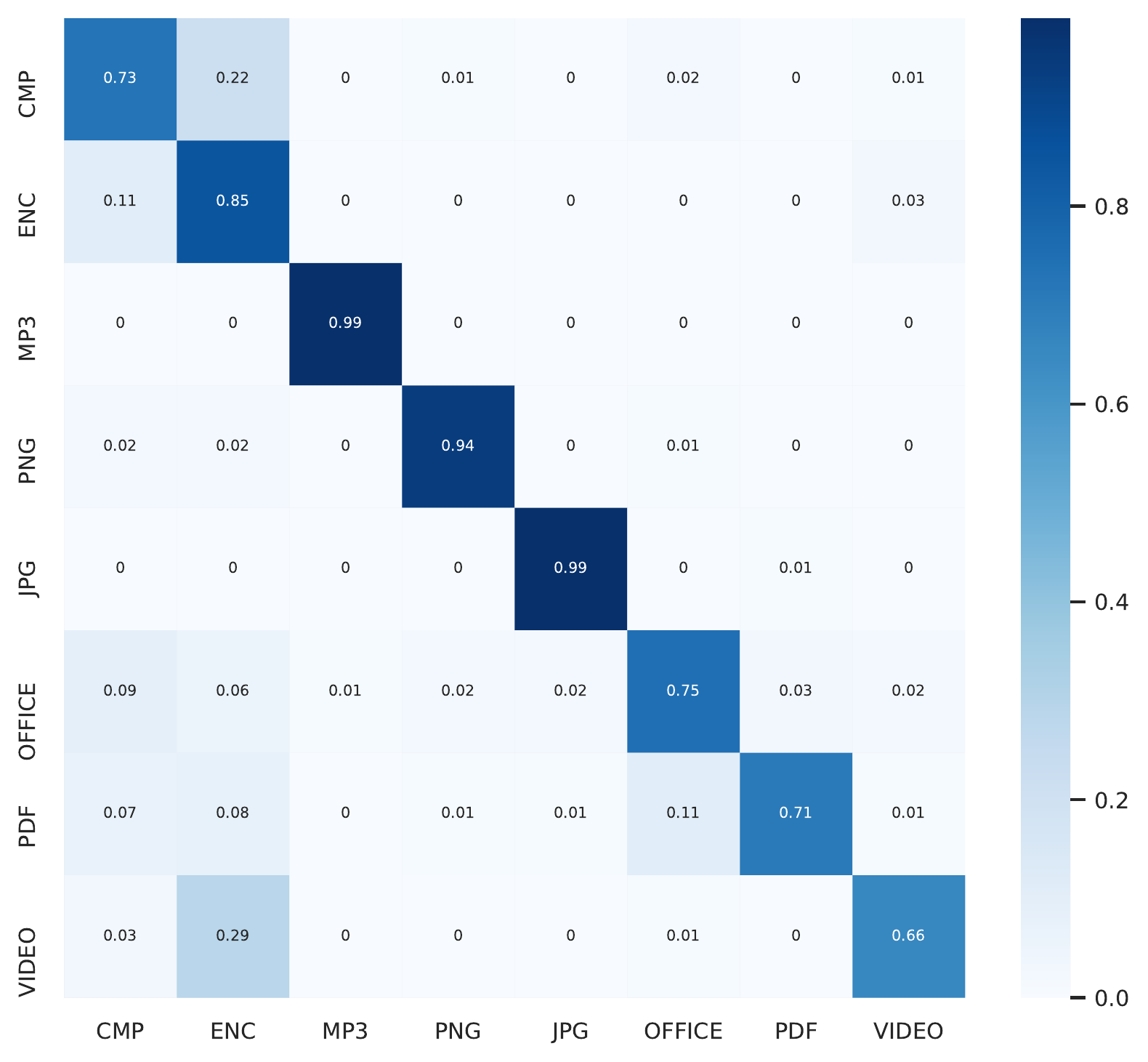}
  \caption{Confusion matrix for the content-type classifier}
  \label{fig:conf_multi_class}
\end{figure}
Our multiclass classifier has the ability to (1) distinguish encrypted and compressed data, and (2) pinpoint the specific format compressed data belong to. This is a significant improvement over the functionality of existing tests, that can only distinguish encryption and compression, but cannot tell the specific format or even generic type of the compressed data. 
In this section, we analyze the effectiveness of our multi-class classifier in fingerprinting the correct type of compressed content. Since for some type families we have many subtypes (e.g., for video we have 5 types, for cmp we have 4), while for others only one or two, in this experiment we group some of the classes in macro labels. Specifically, the office macro label includes all office file types, the compressed (cmp) macro labels includes zip, gzip, rar, bz2 and the video macro label includes h264, h265, mpeg2, mpeg4 and vp8. 

\begin{figure*}[t]
  \centering
  \includegraphics[width=\textwidth]{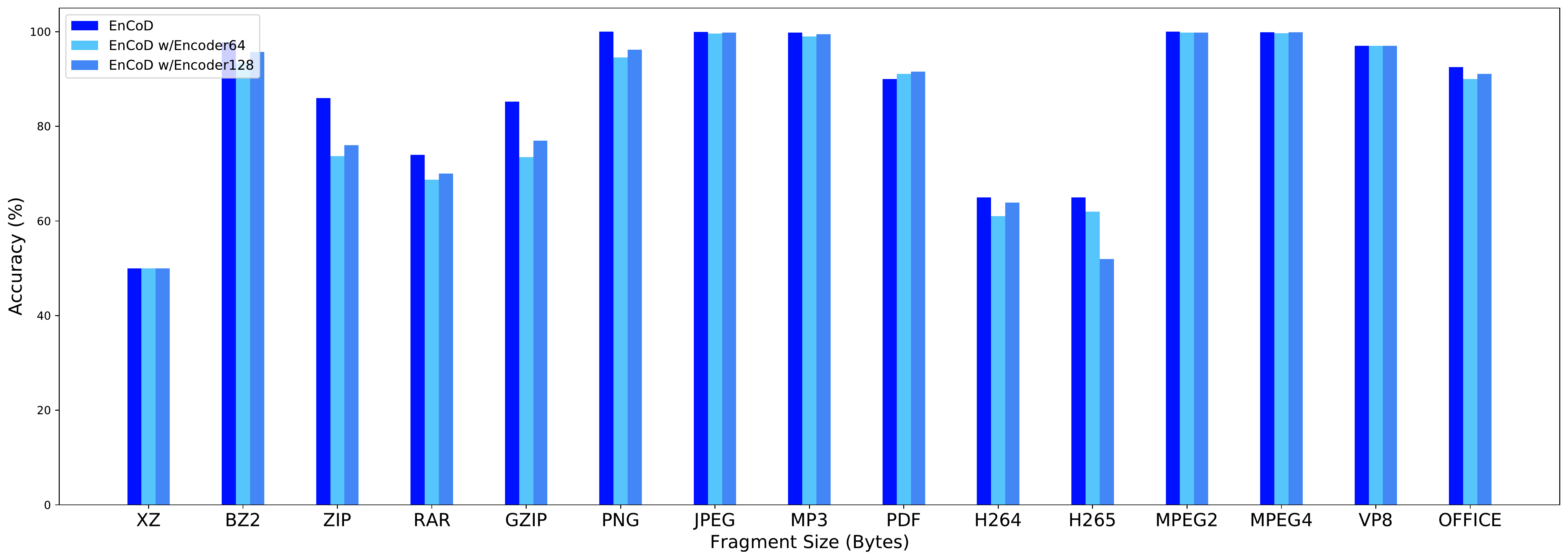}
  \caption{Performance comparison between plain EnCoD binary classifiers, EnCoD binary classifier with Encoder64 and EnCoD binary classifier with Encoder128. Encoder64 uses 64 dimensions for the latent space representation and Encoder128 uses 128 dimensions. All results are for 2048 Bytes data fragments.}
  \label{fig:enc_plain_bar}
\end{figure*}

\begin{figure}[t]
  \centering
  \includegraphics[width=\linewidth]{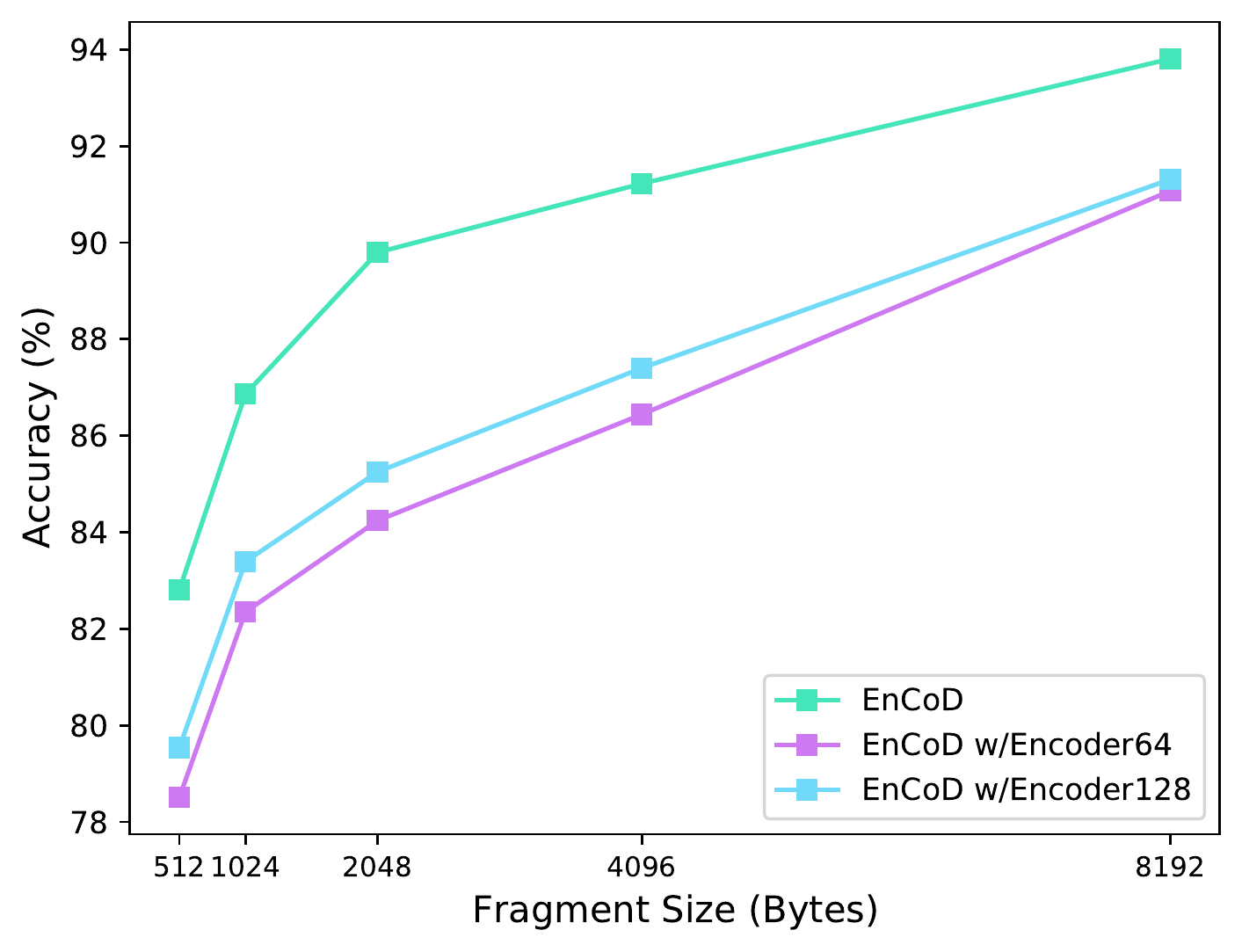}
  \caption{Performance comparison for all formats binary classification between plain EnCoD, EnCoD with Encoder64 and EnCoD  with Encoder128. Encoder64 uses 64 dimensions for the latent space representation and Encoder128 uses 128 dimensions. }
  \label{fig:enc_all_format_bin}
\end{figure}

Figure~\ref{fig:conf_multi_class} shows the confusion matrix for the multi-class classifier. Results indicate that our classifier is able to pinpoint the type of a given sample with consistently high precision for most formats, especially png, mp3, jpeg and encryption. It performs fairly well on the other considered compressed formats such as cmp (which contains a mixture of zip, rar, gzip and bz2 feature vectors) but with a slightly higher rate of misclassified instances between enc and cmp. This can be explained by the fact that their distributions are very close, and intrinsically hard to distinguish. The worse-performing class is the video macro label, for which a sizable portion of the samples is classified as encrypted content. However, given the results of the binary classification presented in Figure~\ref{fig:bcbf}, this behavior is expected. The binary classifiers were already struggling in distinguishing the h264 and h265 types from encryption. Therefore, it is expected that the multiclass classifier would struggle as well with these formats.

\subsection{Autoencoder Approach}
\label{sec:autoenc_eval}
We analyze the effectiveness of the autoencoder approach presented in Section~\ref{sec:autoencoder} compared to the base \textsc{EnCoD} binary approach. As previously discussed, we have two separate encoder architectures, encoder64 and encoder128, compressing the feature vector in a 64-dimensional and 128-dimensional latent representation respectively. This compressed representation is then fed to a binary NN classifier to distinguish between encrypted data and compressed data of a specific type.

Figure~\ref{fig:enc_plain_bar} compares the performance of base \textsc{EnCoD} binary classifiers to the binary classifiers extended with encoder64 and with encoder128. As we can see, the base \textsc{EnCoD} models outperform the encoder models in all binary classification tasks, with the only exception being the pdf type where the encoders slightly outperform the base model. The difference in accuracy is especially apparent for the four general purpose compression formats bz2, zip, rar and gzip, where the best autoencoder (encoder128) results in accuracies 5 to 10 percentage points lower than the base model. Similar results can be observed in Figure~\ref{fig:enc_all_format_bin}, which compares the average accuracy of all binary models for different data fragment sizes. The performance of the encoder-based approaches trails that of the base \textsc{EnCoD} models for all fragment sizes, with accuracy differences in the range of 4 to 10 percentage points.

These results indicate that, while the encoder is able capture meaningful features for classification in the latent representation, it also loses critical information that the plain NN can use to achieve better performance.

\subsection{Overhead}
\label{sec:eval_overhead}In the final part of our evaluation, we analyze the practical applicability of the three approaches, comparing their runtime in order to understand if they can be deployed in time-critical applications. For this test, we used a small dataset comprised of 1000 randomly-selected compressed or encrypted samples. We ran three approaches (NIST, HEDGE and our binary ML model) on each sample, taking  individual runtime and repeating the experiment 1000 times. We did not include the autoencoder approach in this evaluation, given its poor performance when compared to the plain NN models. Table~\ref{tab:overhead} presents the results of our evaluation. As we can see, while both mean and median runtime for NIST tests are faster then HEDGE, our proposed binary classifier is considerably faster than both. Both mean and median runtime for the ML model are three orders of magnitude faster than both NIST and HEDGE, making it easily applicable to scenarios that require fast classification results such as ransomware detection. It is worth noting that the evaluation of our ML model was carried out by measuring the time required to predict a single sample, rather than a batch of samples. However, our model can easily classify multiple samples in parallel by exploiting the heavy parallelism of GPUs, further decreasing the runtime required per individual sample.

\begin{table}[t]
\small
\centering
\begin{tabular}{|l|c|c|c|}  \hline
 \textbf{Approach}      & \textbf{Mean} & \textbf{Median}   & \textbf{Std.dev}  \\\hline
    NIST                &  0.1          &  0.1              & 0.004         \\\hline
    HEDGE               &  0.44         &  0.43             &   0.008       \\\hline
    Binary Classifier   &  0.00046      &  0.00044          &  0.00012        \\\hline
\end{tabular}
\caption{Time required by each approach to classify one sample, in seconds.}
\label{tab:overhead}
\end{table}

\section{Discussion of Findings}
\label{sec:Discussion}

Results shown in Section~\ref{sec:Evaluation} highlight the difficulty of discriminating compressed and encrypted fragments. State-of-the-art statistical tests tend to fare better than entropy measures (ref. Section~\ref{sec:Background}), but their performance varies significantly depending on the specifics of the compressed format and fragment size. Moreover, such approaches can only determine whether a given fragment is encrypted with a certain confidence, but cannot distinguish between different compressed formats. \textsc{EnCoD}, the learning-based approach introduced in Section~\ref{sec:Classifier}, tackles both these limitations. Both per-format and multi-class classifiers outperform existing tests on all considered file types/block sizes. Moreover, our multi-class classifier can be used to determine the format of a given unknown fragment, even in the complete absence of any context or information on its type.

Results show that accuracy improves consistently with increasing fragment size. This is in a sense to be expected; all approaches considered in this paper leverage differences between the byte value distribution of random data (which is uniform) and that of compressed data. Perfectly estimating the byte value distribution of a short data stream is generally not possible. As sequences get shorter, the probability that the estimated distribution may not reflect the typical distribution for their content type increases. However, as the size of the sample increases, the estimated empirical distribution approaches the underlying data distribution, allowing us to capture any deviation from the uniform distribution. For modern compression algorithms, these deviations are quite minor, and a 512-byte block gives even accurate tests very little data to work with. However, when enough data is available, it is possible to identify the class of data with high accuracy; our learning-based classifier exceeds 90\% accuracy already for 2048-byte blocks. In general, we recommend against using any one approach as the sole guidance for automated security decisions (e.g. dropping/allowing flows, terminating processes, etc.). However, when integrated as part of a more complete set of features in a larger system, our proposed classifiers can provide an additional robust feature to use in the decision-making process.

Given the discussion above, we suspect an intrinsic bound on the accuracy reachable by any classifier which looks purely at byte value distributions. However, approaches attempting to parse fragments or identify recognizable structures are likely to incur an impractical computational cost. Moreover, it is not apparent that any such structure is preserved for very short fragment sizes.

\section{Related Work}
\label{sec:RelatedWork}

\subsection{Entropy-based encryption detection}

Use of entropy estimation to detect encrypted content is common in ransomware detection. Proposals such as RWGuard~\cite{mehnaz_rwguard}, UNVEIL~\cite{kirda_unveil}, Redemption~\cite{kirda_redemption} and ShieldFS~\cite{continella_shieldfs:_2016} use entropy of written content either directly as a feature, or as part of feature calculation. It should be noted that none of these detectors use entropy as the sole feature for detection. However, evidence from Section~\ref{sec:Background} suggests that they may be better ignoring entropy altogether.
In the realm of digital forensics, entropy estimation has been used to determine the type of unknown disk data fragments. One of the most complete approaches is that of Conti et al.~\cite{conti_automated_2010}. However, the same authors found that such estimates have limited discerning power in distinguishing encrypted and compressed content, and aggregated the two types under a single label.

Entropy estimation has also been applied to the real-time analysis of network traffic. Dorfinger's Master thesis~\cite{dorfinger_real-time_2010} proposes a system for discriminating encrypted and non-encrypted traffic, to ensure that all communications from a target network are encrypted. Similar approaches were also proposed by Mamun et al.~\cite{mamun_entropy_2016} and Malhotra~\cite{malhotra_detection_2007}. Zhang et al. proposed an entropy-based classifier for the identification of botnet traffic~\cite{zhang_detecting_2013}. All these approaches also suffer from the limitations of using high entropy as a fingerprint of encryption.
Wang et al.~\cite{wang_using_2011} report positive results in using an SVM classifier to discriminate between various data types using entropy estimates. Their application scenario is different from ours, as they consider both low-entropy (non-compressed) and high-entropy (compressed or encrypted) formats. We only consider high-entropy formats, which are difficult to distinguish using entropy alone.

Finally, MovieStealer~\cite{wang_steal_2013} aims at identifying encrypted and decrypted-but-compressed media buffers in order to break DRM. It uses an entropy test to single out encrypted and compressed buffers from other data, and the $\chi^2$-test to distinguish them. It requires 800KB of data to reliably identify random data, which is far beyond the fragment size in the scenarios that we consider.

\subsection{Non-entropy-based approaches}

HEDGE, by Casino et al.~\cite{casino_hedge_2019}, evaluates a combination of $\chi^2$-test and a subset of NIST SP800-22~\cite{rukhin_statistical_2010} to discriminate encrypted and compressed traffic. They use a dataset which is significantly smaller than ours, and do not discuss learning-based approaches. A limitation of this class of approaches is the fairly low accuracy, especially for small block sizes (ref. Section~\ref{sec:Evaluation}). Also, this and other similar approaches based on statistical randomness tests (e.g.,~\cite{lipmaa_data_2017,choudhury_empirical_2020}) cannot distinguish between different types of compressed archives. Mbol et al.~\cite{foresti_efficient_2016} investigate the use of the Kullback-Leibler divergence (relative entropy) to differentiate encrypted files from JPEG images. Their analysis does not investigate other formats, and assumes the availability of blocks of significant size (128 to 512KB) from the beginning of each file. Especially in forensic and networking applications, uninterrupted blocks of such size are difficult to obtain.

While the application of neural networks to the problem at hand is fairly new, there exist some preliminary work. Ameeno et al.~\cite{ameeno_using_2019} show promising preliminary results, however the analysis is limited in scope: it only attempts to distinguish zip archives from rc4-encrypted data, and considers whole files (not fragments). Hahn et al.~\cite{hahn_detecting_2018} perform an exploratory analysis of machine learning models. Their dataset is order of magnitudes smaller than ours, and they lack a comparative analysis of statistical approaches.

\section{Conclusions}
\label{sec:Conclusions}

Discriminating encrypted from non-encrypted content is important for a variety of security applications, and oftentimes tackled via entropy estimation. We comprehensively highlighted the limits of this technique and reviewed the effectiveness of the leading alternative approaches: $\chi^2$-test, NIST SP800-22 test suite, and HEDGE. In addition, we proposed \textsc{EnCoD}, a novel neural network classifier of our own design. In order to ensure generality of results, we created a dataset of 400M fragments covering 5 different sizes and 16 data formats.

Results show that previous state-of-the-art methods have blind spots which result in low accuracy for certain fragment sizes/data types. However, our neural network-based approach appears promising. While statistical tests only discriminate between compressed and encrypted data fragments, our multiclass classifier is able to classify between specific compressed formats with $\sim83\%$ accuracy already on 2KB fragments, and our binary classifier reaches $\sim90\%$ accuracy on the same fragment size. This suggests that systems incorporating encrypted content detection (e.g., ransomware detectors) would be better served by learning-based, rather than hand-crafted statistical approaches. This finding also suggests that learning may have useful applications to other problems in content type inference. Overall, we believe this work is an important step forward towards reliable encryption detection.

\section{Acknowledgments}
We would like to thank Daniele Venturi and Guinevere Gilman for their useful insights and comments.
This work was supported by Gen4olive, a project that has received funding from the European Union’s Horizon 2020 research and innovation programme under grant agreement No. 101000427, and in part by the Italian MIUR through the Dipartimento di Informatica, Sapienza University of Rome, under Grant Dipartimenti di eccellenza 2018–2022.

\bibliographystyle{IEEEtranS}
\bibliography{bibliography}
\begin{IEEEbiography}[{\includegraphics[width=1in,height=1.25in,clip,keepaspectratio]{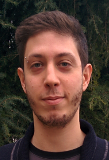}}]{Fabio De Gaspari} is a PostDoc researcher at the Computer Science Department of Sapienza University of Rome, Italy. He received the PhD in Computer Science from Sapienza University in 2019. His research areas are machine learning and security, new network architectures and security and active defence techniques. Fabio  obtained his  MSc  in  Computer  Science with honors in 2015 from the University of Padova, Italy. 
\end{IEEEbiography}

\begin{IEEEbiography}[{\includegraphics[width=1in,height=1.25in,clip,keepaspectratio]{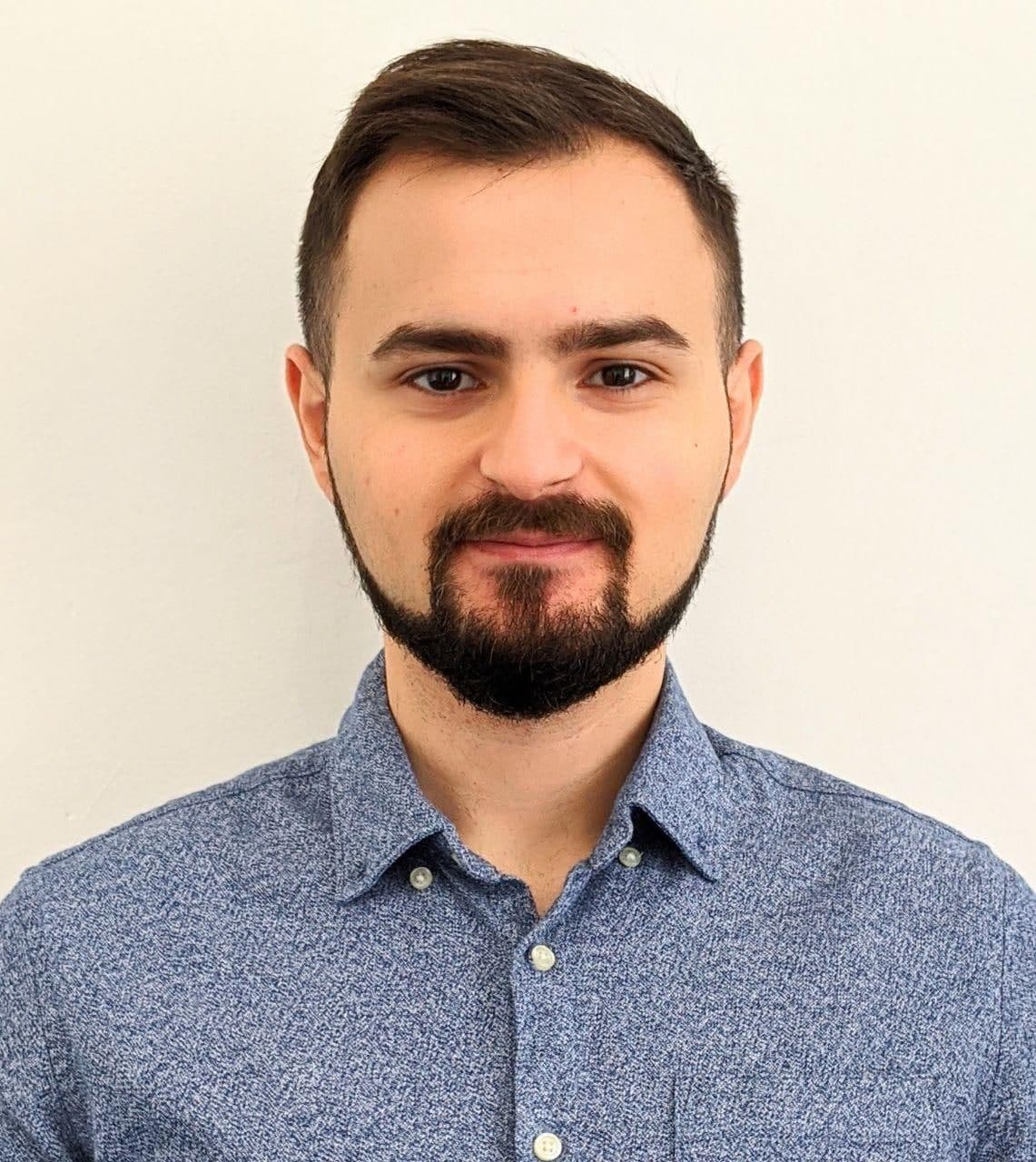}}]{Dorjan Hitaj}
is a PhD student in Computer Science with Sapienza University of Rome. His research focus is on machine learning and security. He obtained his MSc. degree in Computer Science with high honors from Sapienza University of Rome in 2018.
\end{IEEEbiography}

\begin{IEEEbiography}[{\includegraphics[width=1in,height=1.25in,clip,keepaspectratio]{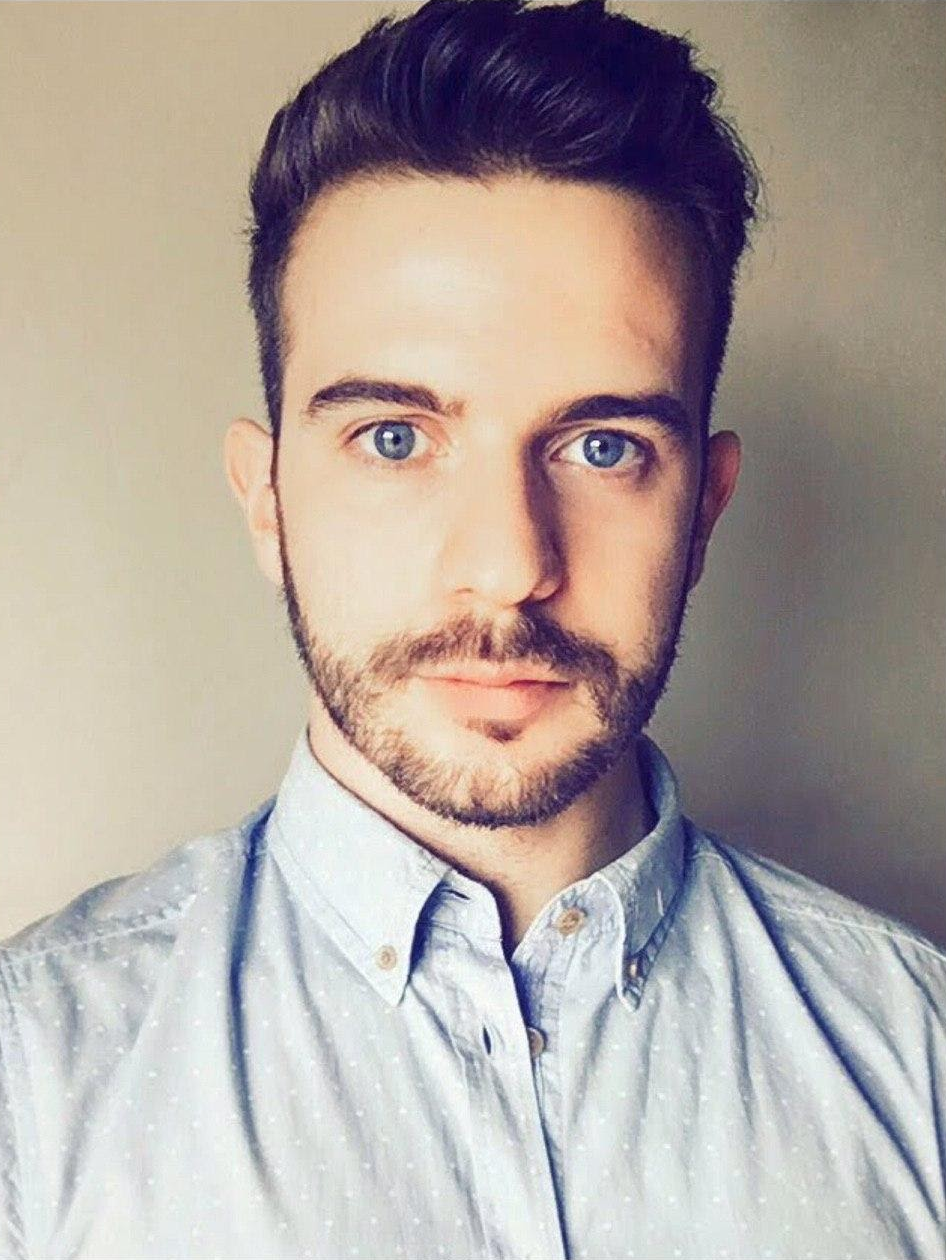}}]{Giulio Pagnotta}
is a Ph.D. student in Computer Science with Sapienza University of Rome. His research focus is on machine learning and security. He obtained his MSc. degree in Cybersecurity with high honors from Sapienza University of Rome in 2019.
\end{IEEEbiography}

\begin{IEEEbiography}[{\includegraphics[width=1in,height=1.25in,clip,keepaspectratio]{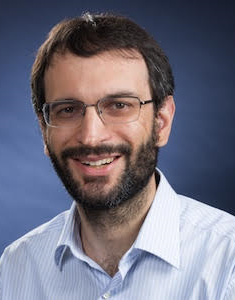}}]{Lorenzo De Carli}’s research interests focus on the security of the web and internet-of-things devices, and usable security. His contributions include hardware accelerators for packet inspection and forwarding, parallelization strategies for intrusion detection, and analysis of malware communications. Lorenzo received a B.Sc. (2004) and a M.Sc. (2007) in Computer Engineering from Politecnico di Torino, Italy, and a M.Sc. (2010) and Ph.D. (2016) in Computer Science from the University of Wisconsin-Madison. 
\end{IEEEbiography}

\begin{IEEEbiography}[{\includegraphics[width=1in,height=1.25in,clip,keepaspectratio]{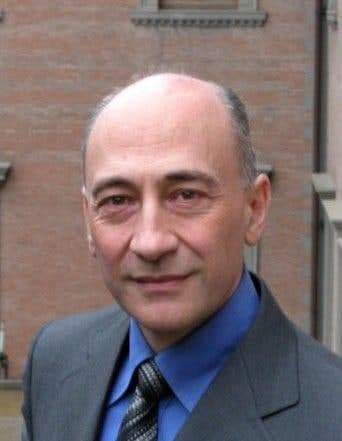}}]{Luigi V. Mancini}~received the PhD degree in Computer Science from the University of Newcastle, U.K., in 1989. He is currently a Full Professor with Sapienza University of Rome, where he is the chair of the Master degree in Cybersecurity. He has authored over 130 scientific papers in international conferences and journals.
\end{IEEEbiography}

\end{document}